\documentclass[runningheads]{llncs}
\usepackage{graphicx}
\usepackage[super]{nth}
\usepackage{booktabs}
\usepackage{multirow}
\usepackage{url}
\usepackage{color}
\usepackage[colorlinks=true, allcolors=blue]{hyperref}
\usepackage{orcidlink}
\usepackage{fontawesome}

\let\savethefootnote\thefootnote
\newcommand\freefootnote[1]{%
  \let\thefootnote\relax%
  \footnotetext{#1}%
  \let\thefootnote\savethefootnote%
}

\begin{document}
\title{Leveraging Self-Supervised Learning for Fetal Cardiac Planes Classification using Ultrasound Scan Videos}
\titlerunning{Leveraging SSL for Fetal Cardiac Planes Classification using US Scan Videos}
%
\author{
Joseph Geo Benjamin\inst{1}\orcidlink{0009-0001-4680-9134}\textsuperscript{(\faEnvelopeO)} \and 
Mothilal Asokan\inst{1}\orcidlink{0009-0005-9461-2749} \and 
Amna Alhosani\inst{1}\orcidlink{0000-0003-4636-8655}\and \\
Hussain Alasmawi\inst{1}\orcidlink{0000-0002-5404-8849} \and 
Werner Gerhard Diehl\inst{2} \and 
Leanne Bricker\inst{2} \and \\
Karthik Nandakumar\inst{1}\orcidlink{0000-0002-6274-9725} \and 
Mohammad Yaqub\inst{1}\orcidlink{0000-0001-6896-1105}
}
\authorrunning{ J. G. Benjamin et al. }
%
\institute{
Mohamed bin Zayed University of Artificial Intelligence, \\ Abu Dhabi, United Arab Emirates \\
\email{\{joseph.benjamin, mothilal.asokan, amna.alhosani, 
hussain.alasmawi, karthik.nandakumar, mohammad.yaqub\}@mbzuai.ac.ae} \and
Abu Dhabi Health Services Company (SEHA), Abu Dhabi, United Arab Emirates \\
\email {\{wernerd, LeanneB\}@seha.ae} \\
}

\maketitle 

\begin{abstract}
Self-supervised learning (SSL) methods are popular since they can address situations with limited annotated data by directly utilising the underlying data distribution. However, adoption of such methods is not explored enough in ultrasound (US) imaging, especially for fetal assessment. 
We investigate the potential of dual-encoder SSL in utilizing unlabelled US video data to improve the performance of challenging downstream Standard Fetal Cardiac Planes~(SFCP) classification using limited labelled 2D US images.
We study $7$ SSL approaches based on reconstruction, contrastive loss, distillation and information theory, and evaluate them extensively on a large private US dataset. Our observations and finding are consolidated from more than $500$ downstream training experiments under different settings.
Our primary observation shows that for SSL training, the variance of the dataset is more crucial than its size because it allows the model to learn generalisable representations which improve the performance of downstream tasks. 
Overall, the BarlowTwins method shows robust performance irrespective of the training settings and data variations, when used as an initialisation for downstream tasks. Notably, full fine-tuning with $1\%$ of labelled data outperforms ImageNet initialisation by $12\%$ in F1-score and outperforms other SSL initialisations by at least $4\%$ in F1-score, thus making it a promising candidate for transfer learning from US video to image data. Our code is available at \url{https://github.com/BioMedIA-MBZUAI/Ultrasound-SSL-FetalCardiacPlanes}.

\end{abstract}

\freefootnote{J.G. Benjamin and M. Asokan --- Contributed equally.}
%
%
%

\section{Introduction}
Fetal sonography is used to assess the growth and well-being of the fetus. The ISUOG\footnote{ International Society of Ultrasound in Obstetrics and Gynecology} guidelines \cite{https://doi.org/10.1002/uog.26224} and the FASP\footnote{Fetal Anomaly Screening Programme NHS UK} handbook \cite{nhsfasp2023} recommend the acquisition and use of standardised planes for fetus abnormality and growth assessment which is done manually by sonographers.
In practice, a well-trained sonographer should account for variations caused by fetal movement \& position, maternal body habitus, probe placement angle, etc. At the device level, even calibration and manufacturing differences can produce variations in image quality and measurements. This makes it hard to acquire Standard Fetal Planes (SFP) consistently and even more complicated for Standard Fetal Cardiac Planes (SFCP) which is critical in assessing conditions such as congenital heart diseases and intrauterine growth restrictions.
Building automated systems to tackle aforementioned issues faces challenges due to large intra-class variations and inter-class similarities among the anatomical structures. This becomes even more challenging for SFCP, with fast motion due to heartbeats, leading to many misclassifications.

A myriad of work exists to solve the automated FSP classification using data-driven approaches like supervised machine learning~\cite{yaqub2015} and deep learning~(DL)~\cite{7974824,7875138} with fetal ultrasound~(US) images. But labelling large amounts of data that can help capture class variability and distribution shifts is expensive. In addition, unlike natural images, the existence of large public datasets is also hindered by privacy concerns. In most healthcare facilities, large volumes of unlabelled data will be found in isolation, which could neither be shared publicly nor be labelled to utilise privately. Recent self-supervised learning (SSL) techniques mitigate the requirement of large labelled datasets to train good DL models. Although SSL methods have been applied on US imaging analysis especially echocardiography~\cite{Holste2022SelfSupervisedLO,MSaeed2022}, it is understudied in the fetal image analysis field. Since US scanning involves the recording of fetal scans as videos alongside the acquisition of 2D images, it can be leveraged for data-hungry self-supervision methods and thus can be utilised on private data available at healthcare facilities to create/improve AI systems. 

In this work, we aim to clarify the following two questions regarding the dual-encoder SSL methods. 
\textit{How does SSL pretraining on US video data impact downstream SFCP classification with limited labelled data?}
\textit{Which SSL method is effective in utilizing US video data?}

We believe that answering these questions will facilitate practical decision-making in a broad scope and easier adoption of leveraging real-world healthcare data instead of relying on complex engineering techniques to achieve good performance.
This work does not intend to provide a new technical addition to the deep learning community. The research contribution of this work is to provide a thorough analysis of a set of well-established SSL methods, that strictly do not require labelled data, for the problem of fetal US image classification. We conduct several ablations for SSL training with different frame sampling, amount of data and seed weights which leads to some interesting implications that are important to disseminate to the research community and help make better use of unlabelled fetal US videos.

\section{Related Work}

SSL methods have been explored for utilizing fetal US videos with pretext tasks such as correcting reordered frames and predicting geometric transformations~\cite{9098666} or restoring altered images~\cite{CHEN2019101539} to learn transferable representations for downstream tasks. 
More recently, SSL has moved towards dual-encoder architectures \cite{bardes2022vicreg,10.5555/3524938.3525087,10.5555/3495724.3497510,9157636,pmlr-v139-zbontar21a}(similar to siamese network) which rely on the distribution of data itself to learn meaningful representations rather than crafting pretext tasks that suit specific problems/data of interest. This line of SSL methods has not gained much focus for applications utilizing US video.
A comprehensive survey by Fiorentino et al.~\cite{FIORENTINO2023102629} studies DL methods in fetal sonography and highlights recent trends and challenges. This shows a gap in the adoption of SSL, especially dual-encoder methods for US videos. Benchmark analysis by Taher et al.~\cite{10.1007/978-3-030-87722-4_1} shows the effective transferability of self-supervised pretraining over supervised pretraining using ImageNet~\cite{5206848} dataset for a variety of medical imaging tasks.

The work by Fu et al.~\cite{10.1007/978-3-031-25066-8_23} incorporates a contrastive SSL approach with anatomical information by utilising labels. Zhang et al.~\cite{10.1007/978-3-031-26351-4_1} proposed hierarchical semantic level alignments for US videos using contrastive learning with labels through a smoothing strategy to improve the transferability. Different from these works, our study focuses on leveraging medical data itself i.e. US scan videos for SSL with no annotation information.
A survey by Schiappa et al.~\cite{10.1145/3577925} provides a detailed review and comparison of SSL techniques including dual-encoder using contrastive methods in the natural video domain.

\section{Methodology}

\subsection{Data and Preprocessing}

We perform our experiments on a large private fetal US scan data. This dataset consists of two modalities, labelled images of SFP and unlabelled videos (mainly SFCP and a few other views) collected from pregnant patients during their second trimester screening. The data is gathered over one calendar year and across different machine types (Voluson E8/E10/P8/S10-Expert/V830).
\begin{table}[tb]
\centering
\caption{Subtable.1 indicates the classwise imbalance both in terms of the images and patients, Subtable.2 shows different sampling frequency and frame count (images) used for SSL training and Subtable.3 shows the statistics of Video. 
}
\label{tab:data}
\begin{tabular}{l p{0.5em}rrrrr p{0.5em}rrrrr}

\toprule[1pt]
\midrule[0.3pt]
\multirow{2}{3em}{Class} && \multicolumn{5}{c}{Images} && \multicolumn{5}{c}{Patients} \\
\cmidrule{3-7} \cmidrule{9-13}
          && Train && Valid && Test && Train && Valid && Test \\
\midrule
3VV/3VT   && 1703  && 170   && 580  && 1013  && 96    && 342  \\
4CH       && 2699  && 307   && 876  && 1317  && 155   && 438  \\
LVOT      && 4371  && 464   && 1439 && 2017  && 228   && 663  \\
RVOT      && 4036  && 442   && 1306 && 1974  && 222   && 650  \\
Non-Heart && 4400  && 462   && 1441 && 2434  && 254   && 754  \\
\midrule
Total     && 17209 && 1845  && 5642 && 3198  && 359   && 1033  \\

\midrule[0.3pt]
\bottomrule[1pt]
\end{tabular}
\hspace{1em}
\begin{tabular}{lc}
\toprule[1pt]
\midrule[0.3pt]
\multirow{2}{4em}{Sampling Freq} & \multirow{2}{4em}{V.Frame Count}  \\
&\\
\midrule
All frames      & 405363  \\
Every $5^{th}$   & 81556  \\
Every $35^{th}$  & 12217  \\
Every $70^{th}$  & 6464  \\
1 per video    & 1349  \\
\midrule\midrule
\textbf{Patients Count: }   &  \textbf{575} \\
\midrule[0.3pt]
\bottomrule[1pt]
\end{tabular}

\vspace{3ex}  
\begin{tabular}{l ccccc ccccc}
\toprule[1pt]
\midrule[0.3pt]
Video Stat.  && mean && std && median && min && max \\
\midrule
Frame Rate  && 70  &&   27 &&  69 && 11  && 123 \\
Frame Count && 456 &&  245 && 358 &&  3  && 800 \\
\midrule[0.3pt]
\bottomrule[1pt]
\end{tabular}

\end{table} 
For classification~(Cls), we use four classes corresponding to the following standard cardiac planes: 3 Vessels View/3 Vessels Trachea view (3VV/3VT), 4 Chamber view (4CH), Left Ventricular Outflow Tract view (LVOT), and Right Ventricular Outflow Tract view (RVOT) and sample few non-heart SFP and create a \nth{5} class corresponding to a non-heart view. Table \ref{tab:data} shows the distribution of images and patients in the dataset.
The datasets are split at the patient level to avoid any information leakage about the classification test set, even patients in the validation/test set were removed from US videos used for SSL training.
\noindent \textit{\textbf{Preprocessing:}} We filter out videos that have Doppler \& split views or any other artifacts. To prevent any shortcut learning, we perform inpainting following the approach described in \cite{9434112} on videos/images thereby removing any inframe marking or annotations done by sonographers. 
We further verify the cleanness of preprocessing by training a ResNet-18 classifier with processed data and applying Grad-CAM \cite{8237336} on a random test subset. we observe that the network focuses on heart features rather than inpainted regions.

\subsection{Self-Supervision Procedure}
To study the benefits of various SSL methods adopted for pretraining, we select methods belonging to different strategies 
\newline\indent\indent \textbf{\textit{(a) Reconstruction}} - AutoEncoder \cite{AE-masci2011stacked}, Inpainting \cite{AEp-Pathak_2016_CVPR} 
\newline\indent\indent \textbf{\textit{(b) Contrastive Loss}} - SimCLR-v2 \cite{10.5555/3524938.3525087}, MoCo-v2 \cite{9157636} 
\newline\indent\indent \textbf{\textit{(c) Distillation-based}} - BYOL \cite{10.5555/3495724.3497510} 
\newline\indent\indent \textbf{\textit{(d) Information theory}} - VICReg \cite{bardes2022vicreg}, BarlowTwins \cite{pmlr-v139-zbontar21a}
\newline
These methods do not explicitly require labelled data which is a critical consideration as we use unlabelled scan videos. We use ResNet-50 as the backbone network along with the appropriate projector network as mentioned in the literature for each dual-encoder method and a convolutional decoder network to output an image plane for reconstruction methods. \\
\noindent \textit{\textbf{Weight initialisation:}}
We study the effect of weight initialisation on SSL training by comparing ImageNet classification pretrained weights and random weights initialisation, both as available in PyTorch. \\
\noindent \textit{\textbf{Hyperparameters:}}
We follow optimal hyperparameters, optimizer settings, and augmentations as suggested in the respective literature of all the dual-encoder SSL methods\cite{bardes2022vicreg,10.5555/3524938.3525087,10.5555/3495724.3497510,9157636,pmlr-v139-zbontar21a}.
We intend to identify the approach that works consistently without dataset specific tweaks or grid searches, as it would be infeasible or compute expensive in many real-world deployments.  
For AutoEncoder and Inpainting training, we use AdamW optimizer with a learning rate of $10^{-3}$, a weight decay of $10^{-6}$, a StepLR scheduler with stepsize $50$, and a gamma $0.5624$. All the methods are trained for 1000 epochs.\\
\textit{\textbf{Batch Size:}} Training SSL with larger batch sizes is known to yield better performance in final downstream tasks. But we use a batch size of $256$ to make fair comparisons under a practical setting because many facilities might not have IT~infrastructure that supports the large batch sizes recommended by the original works. The chosen size could be fit in a single NVIDIA A100-SXM4-40GB machine without memory overflows for SSL training. \\
\noindent \textit{\textbf{Video Frame Sampling Frequency:}}
We conduct experiments using data created by sampling every \nth{5} frame from each video by default and to study the effect of sampling, we conduct a separate experiment with varying sampling frequency for SSL training as shown in Table \ref{tab:data}. Though sampled at a fixed frequency, the difference in frame rate and frame count in each video produces the effect of sampling at different time intervals for each video ensuring variance in data distribution. 
Irrespective of sampling frequency, \nth{1} frame of a video is always included in training data. This is to make sure that at least one frame of each video is included in SSL training even with a larger sampling frequency.

\begin{figure}[t]
    \includegraphics[width=\textwidth]{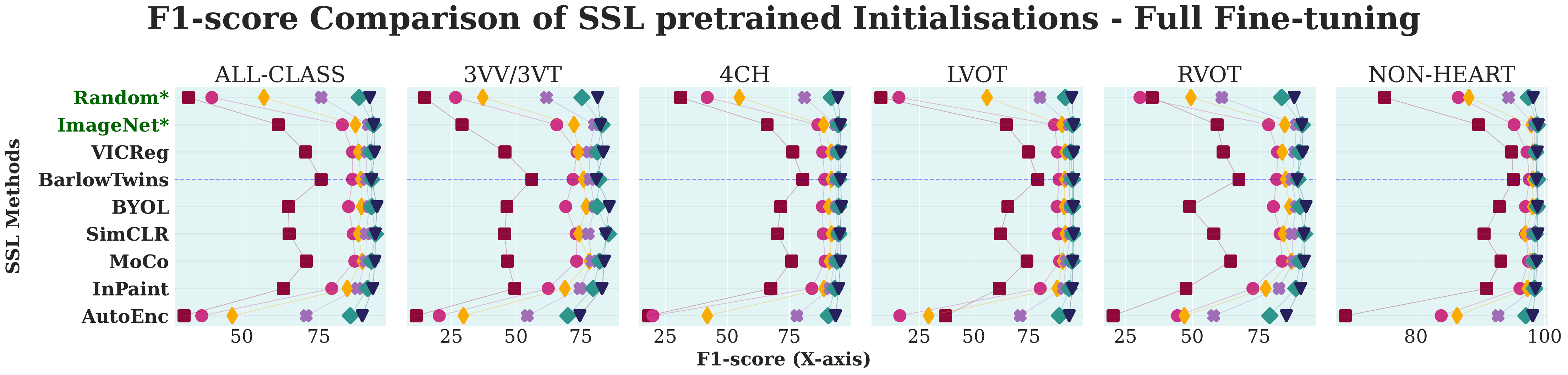}
    
    \includegraphics[width=\textwidth]{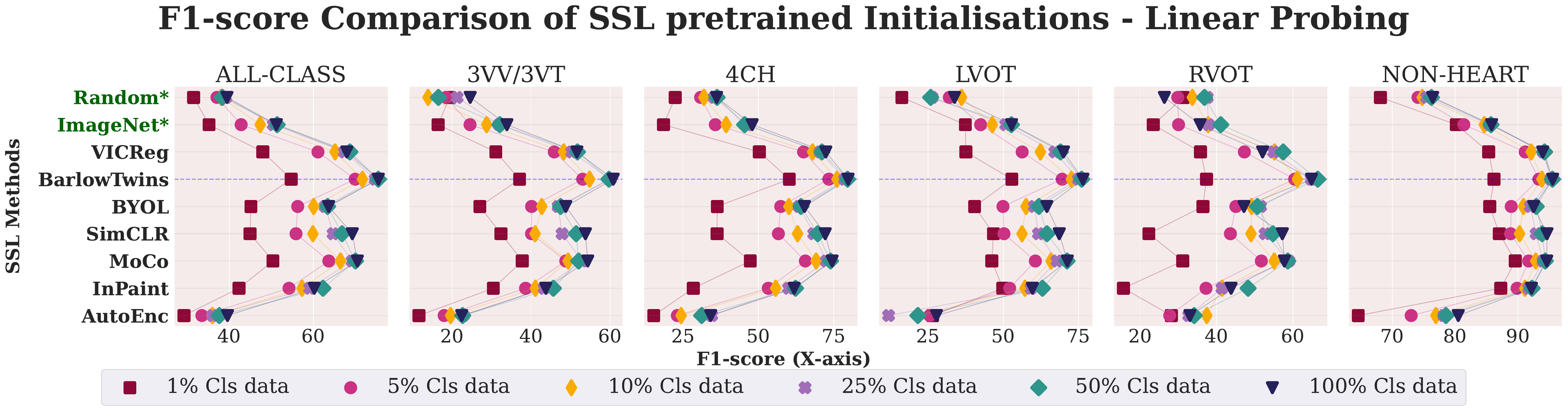}
    \caption{ BarlowTwins performs consistently better even for challenging views. $*$ indicates Non-SSL initilisations.} 
    \label{fig:ssl-main-f1}
\end{figure}

\subsection{Classification Procedure}
We use a network initialised with different SSL pretrained weights \textit{(SSL-weight)} as the feature extractor and attach a linear classifier layer on top to train for downstream tasks. We perform full network fine-tuning to gauge the adaptability of pretrained weights to the downstream task.
We also perform linear probing to understand the linear separable quality of the representations learned during SSL training. We freeze the entire backbone network, attach the BatchNorm layer with ($\gamma=1$ $\beta=0$) and fine-tune only the linear classifier layer. 
Along with \textit{SSL-weights}, we run classification training with random (Kaiming) and ImageNet pretrained weights for comparison.\\
\noindent \textit{\textbf{Hyperparameters:}}
We use AdamW optimiser with a learning rate of $10^{-3}$, a weight decay of $10^{-6}$ without any scheduler, and a batch size of 128. We run the experiments for 100 epochs and select the model at an epoch with the best F1-score in the validation set. \\
\noindent \textit{\textbf{Labelled Data Size:}}
From the entire ($100\%$) classifier training data we obtain $50\%, 25\%, 10\%, 5\%, 1\%$ of data using a stratified sampling technique and run classification experiments on each of them separately. The sample images in each split are kept the same for all the experiments. The F1-score is reported for a fixed number of test samples. This setup enables us to understand the data efficiency achieved by different SSL methods.

\section{Results and Discussions}

\begin{figure}[t]
    \includegraphics[width=\textwidth]{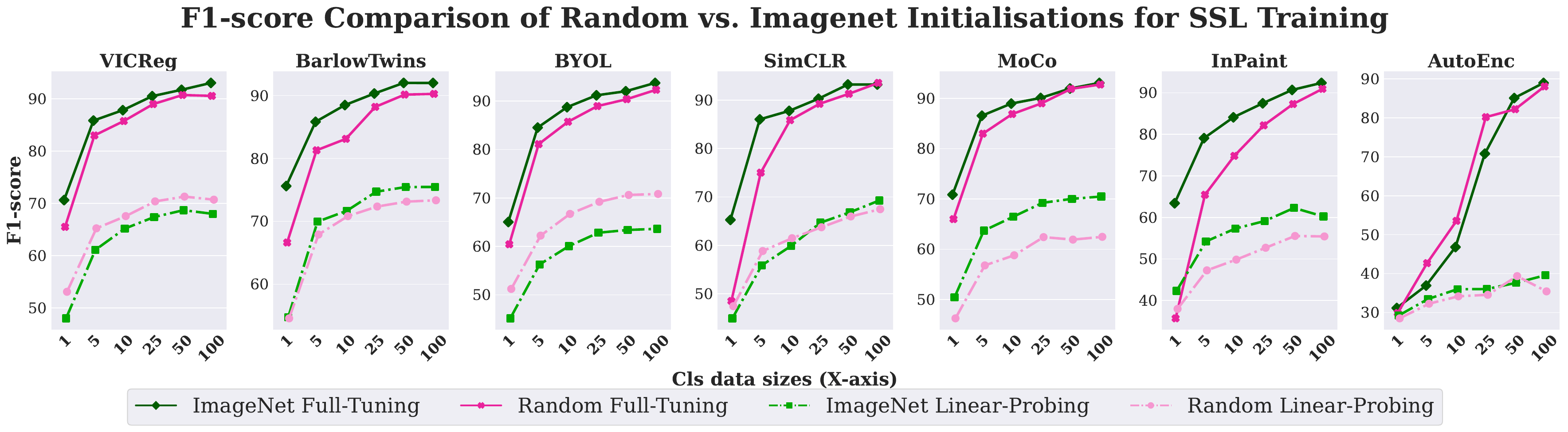}
    \caption{Linear probing shows a different trend than full fine-tuning in random $vs.$ Imagenet initialisation for some SSL training.
    } 
    \label{fig:ssl-imgnet-rand}
\end{figure}

\subsubsection{\textit{How do SSL pretrained models perform on different data sizes?}}
Under full fine-tuning, the SSL pretrained weights \textit{(SSL-weight)} perform better than the de facto ImageNet initialisation when the annotated data size is low. But as the annotated data size increases, the gains diminish and for $100\%$ of the data to fine-tune on, the difference becomes marginal. Even randomly initialised weights for classification show comparable results in larger data setting. F1-Scores for different \textit{SSL-weights} across different data sizes are shown in Figure~\ref{fig:ssl-main-f1}.

Since we train SSL on video data and train the downstream classification on a different set of 2D image data, the aforementioned observation could be because of the following reasons: (a) the data available for SSL training is not very representative of the entire distribution which can lead to limited generalisation, or (b) gain of transfer learning diminishes as the amount of labelled data is more \cite{8954384} although it might help in faster convergence of the models. 
Generative methods perform poorly compared to other SSL methods, notoriously AutoEncoders only learn to memorize the input and reconstruct without learning any contextual information. Amongst the SSL methods, BarlowTwins gives a significant gain performance followed by MoCo and VICReg. 
It is observed that these methods that reduce the contrastive loss or maximize the statistical variance within a batch, underperform BarlowTwins which only decorrelates the representation space. We conduct linear probing of \textit{SSL-weights} to understand the quality of representations learned across models and classes. We observe that outcomes of BarlowTwins followed by MoCo and VICReg are consistently better.

\begin{figure}[t]
    \includegraphics[width=\textwidth]{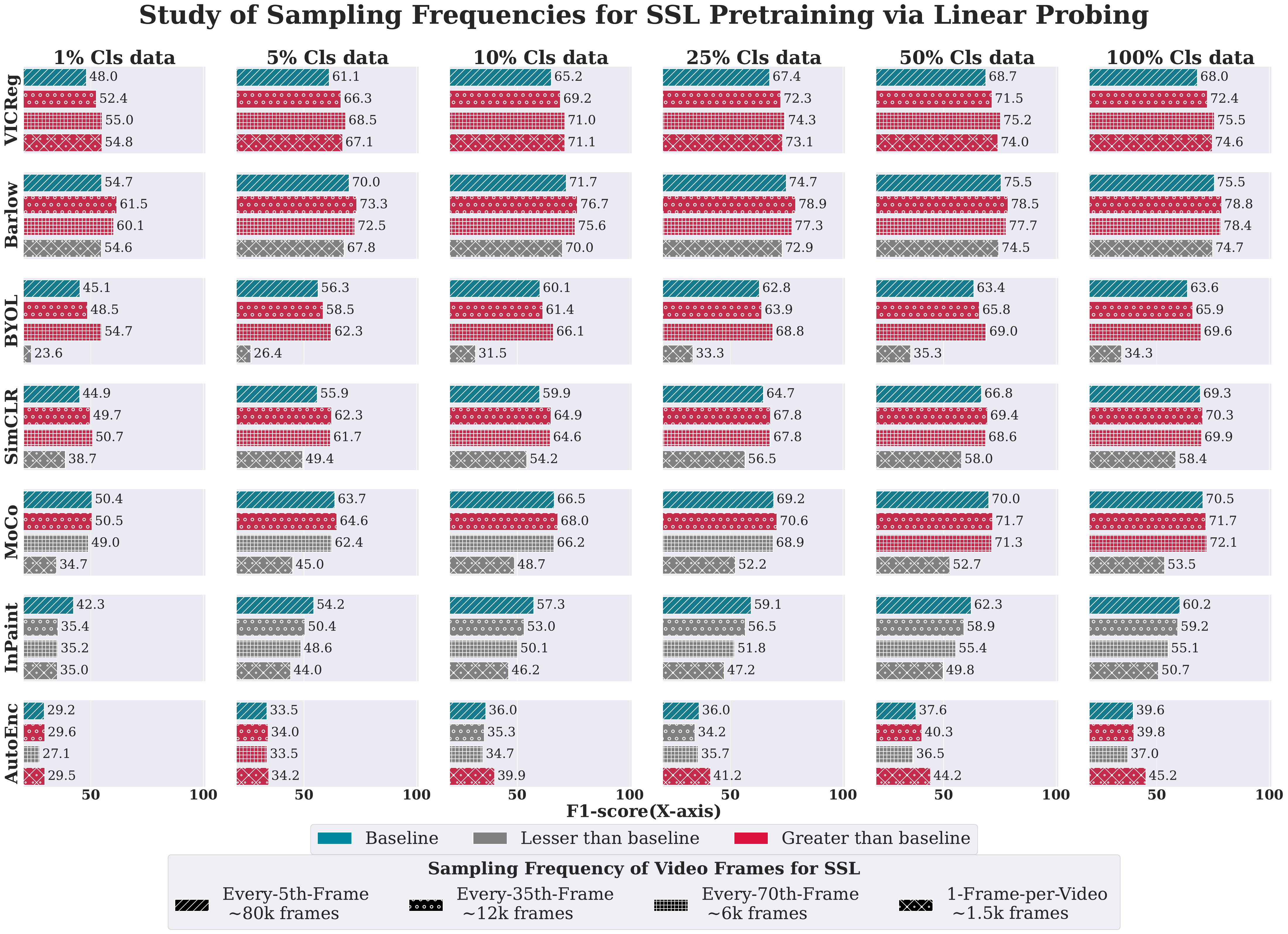}
    \caption{
    Results show trade-off between data variance $vs.$ data size for SSL trainings.
    }
    \label{fig:ssl-volumes-f1}
\end{figure}

\subsubsection{\textit{What is the effect of Random $vs.$ Imagenet initialisation during SSL training?}}
We observe that ImageNet weight initialisation at the beginning of SSL training \textit{(Imnet-setting)} yields a noticeable gain in accuracy during the full fine-tuning of the downstream task compared to the random initialization \textit{(Rand-setting)}. The results are compared in Figure \ref{fig:ssl-imgnet-rand}. 
We concur that ImageNet initialisation gives better generalisation capability by converging weights to a better representational function during SSL training.
Surprisingly, when evaluated with linear probing, we observe \textit{Rand-setting} outperforms \textit{Imnet-setting} for SSL methods such as BYOL, VICReg and marginally in SimCLR.  This indicates that representations learnt by these methods under \textit{Rand-setting} are inherently better than \textit{Imnet-setting}. Yet for the same SSL methods during full fine-tuning \textit{Imnet-setting} is better. 
We reason that certain inductive biases encoded in Imagenet weights that help in generalization, might not be sufficiently adapted for the US dataset during the SSL training phase of these methods. Whereas in \textit{Rand-setting} model has to learn US data-specific cues during SSL training to converge from a random state. Thus, \textit{Imnet-setting} performs poorly in linear probing. But inductive biases kick in during full fine-tuning of \textit{Imnet-setting}, aiding in generalization which leads to better results. 
Interestingly BarlowTwins and MoCo consistently perform better in both full fine-tuning and linear probing, which could mean that they leverage ImageNet specific biases effectively during SSL training itself. This could also be the reason for relatively superior performance compared to other methods that follow a similar SSL training strategy (BarlowTwins $vs.$ VICReg and MoCo $vs.$ SimCLR/BYOL).

\begin{figure}[t]
    \includegraphics[width=\textwidth]{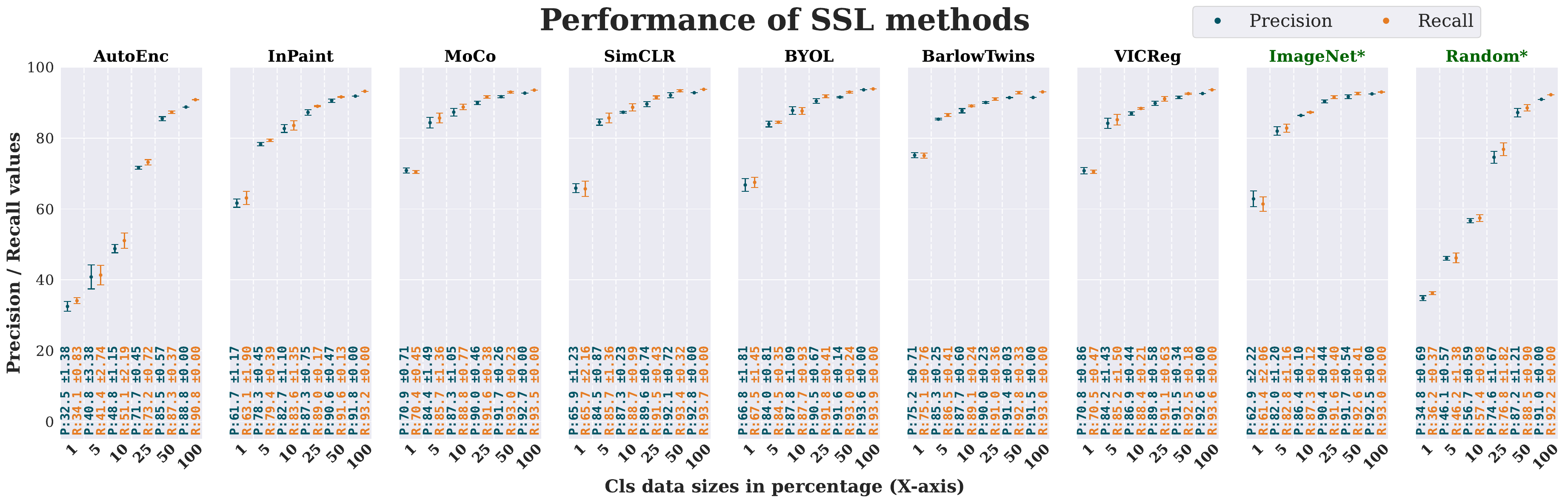}
    \caption{
    Mean \& SD obtained by training with $3$ different sampling of labelled data and seed values.
    } 
    \label{fig:ssl-mean-std}
\end{figure}

\subsubsection{\textit{Does sampling more frames from Videos help improve SSL training?}}
We train \textit{SSL-weight} with different video frame sampling frequencies, such that a higher sampling frequency leads to a lower frame count for training. We conduct linear probing to understand the quality of representations learned. We make an interesting observation in Figure \ref{fig:ssl-volumes-f1} that for many cases, as we use lesser frames per video (high sampling frequency) for SSL  pretraining, the accuracy increases. 
This might be counterintuitive to the generally held notion that a larger dataset can enhance SSL performance. Though the number of frames per video increases, the variance of samples in a batch throughout the training decreases. 
As many of these SSL methods directly or indirectly rely on batch variance for learning good representations \cite{shwartz-ziv2022what}, batches with lesser variance seem to impact learning.  In such cases, highly redundant mutual information also hurts SSL training \cite{10.5555/3495724.3496297}.
But this trend breaks as soon as SSL data size decreases drastically, indicating that there should be an ideal balance between the amount of data and its variance to achieve better performance.
The influence of data distribution on learning varies across different methods for e.g. VICReg is the most dependent on variance than the size of data while the Inpainting method is least dependent (although overall performance is poor). 
Figure \ref{fig:ssl-mean-std} shows precision and recall values.

\section{Conclusion}
In this work, we conduct extensive experimentation to understand the behaviour of various SSL methods in utilising fetal US scan videos. Specifically, we study their empirical value in Cardiac Planes (SFCP) classification under real-world medical constraints. Our observations show that SSL methods give a boost in performance under limited annotated data. We found that BarlowTwins is most robust to variations in data distribution/size and training settings and gives consistent performance. In the scope of this study, we do not consider different backbones or methods that leverage label information during SSL training, since our motive is to evaluate the utility of SSL methods requiring no labels. However, our findings could be further extended with different backbones or methods leveraging labels during SSL training. We believe that our findings will lay a firm foundation for future works focused on recent forms of SSL methods for the US domain, especially in leveraging video data.

%
%
%
\bibliographystyle{splncs04}
\bibliography{references}

\end{document}